\documentclass[aoas,nameyear,rotating,dvips]{arximspdf}
\usepackage{dcolumn,multirow}
\usepackage{graphicx}

\doi{10.1214/09-AOAS264}
\volume{3}
\issue{4}
\pubyear{2009}
\firstpage{1521}
\lastpage{1541}

\makeatletter
\newcolumntype{d}[1]{D{.}{.}{#1}}

\DeclareMathAlphabet\mathcaligr{OMS}{cmsy}{m}{n}
\makeatother

\begin{document}
\begin{frontmatter}

\title{A branching process model for flow cytometry\\ and
budding index measurements in cell\\ synchrony experiments}
\runtitle{Branching process model for cell synchrony experiments}

\begin{aug}
\author[A]{\fnms{David A.} \snm{Orlando}\thanksref{t3,t2}\ead[label=e1]{david.orlando@duke.edu}},
\author[B]{\fnms{Edwin S.} \snm{Iversen Jr.}\corref{}\thanksref{t3}\ead[label=e2]{iversen@stat.duke.edu}},\break
\author[C]{\fnms{Alexander J.} \snm{Hartemink}\thanksref{t1}\ead[label=e3]{amink@cs.duke.edu}}\and
\author[D]{\fnms{Steven B.} \snm{Haase}\thanksref{t2}\ead[label=e4]{shaase@duke.edu}}
\runauthor{Orlando, Iversen, Hartemink and Haase}
\affiliation{Duke University}
\thankstext{t3}{These authors contributed equally to this work.}
\thankstext{t1}{Supported in part by the Alfred P. Sloan Foundation,
National Science Foundation, and National Institutes of Health.}
\thankstext{t2}{Supported in part by the American Cancer Society and
National Institutes of Health.}
\address[A]{D. A.~Orlando \\
Program in Computational Biology\\
\quad\& Bioinformatics\\
102 North Building\\
Box 90090\\ Duke University\\
Durham, North Carolina 27708\\ USA\\
\printead{e1}}
\address[B]{E. S.~Iversen Jr.\\
Department of Statistical Science\\
Box 90251\\
Duke University \\
Durham, North Carolina 27708-0251 \\
USA\\ \printead{e2}}
\address[C]{A. J.~Hartemink\\
Department of Computer Science\\
Box 90129\\
Duke University \\
Durham, North Carolina 27708\\
USA\\
\printead{e3}}
\address[D]{S. B.~Haase\\
Department of Biology\\
Duke University \\
Durham, North Carolina 27708-1000\\ USA\\
\printead{e4}}

\end{aug}

\received{\smonth{12} \syear{2008}}
\revised{\smonth{4} \syear{2009}}

%
\begin{abstract}
We present a flexible branching process model for cell population
dynamics in synchrony/time-series experiments used to study important
cellular processes. Its formulation is constructive, based on an
accounting of the unique cohorts in the population as they arise and
evolve over time, allowing it to be written in closed form. The model
can attribute effects to subsets of the population, providing
flexibility not available using the models historically applied to
these populations. It provides a tool for \textit{in silico}
synchronization of the population and can be used to deconvolve
population-level experimental measurements, such as temporal
expression profiles. It also allows for the direct comparison of assay
measurements made from multiple experiments. The model can be fit
either to budding index or DNA content measurements, or both, and is
easily adaptable to new forms of data. The ability to use DNA content data
makes the model applicable to almost any organism. We describe the
model and illustrate its utility and flexibility in a study of
cell cycle progression in the yeast \textit{Saccharomyces cerevisiae}.
\end{abstract}

%
\end{frontmatter}

\section{Introduction}\label{sec:intro}

In this paper we describe a novel branching process model that
characterizes the temporal evolution of population heterogeneity in
cell synchrony experiments. These experiments are designed to measure
the dynamics of fundamental biological processes related to the cell's
progression through the cell division cycle. Careful characterization
of these dynamic processes requires experiments where quantitative
measurements are made over time. In many cases, accurate measurements
cannot be made on single cells because the quantitative methods lack
the sensitivity to detect small numbers of biomolecules. For example,
accurate quantitative measurements of genome-wide transcript levels by
microarray require more mRNA than is physically available within a
single cell. Thus, researchers are forced to work with populations of
cells that have been synchronized to a discrete cell cycle state.

Two distinct problems arise in these synchrony/time-series experiments.
First, synchronized populations are never completely synchronous to
begin with, and tend to lose synchrony over time. The lack of perfect
synchrony at any given time leads to a convolution of the measurements
that reflects the distribution of cells over different cell cycle
states. Second, multiple synchrony experiments are often needed to
measure different aspects of a process, and it is often desirable to
compare the temporal dynamics of these aspects. However,
synchrony/time-series experiments, even in the best of experimental
circumstances, exhibit considerable variability which make time-point
to time-point, cross-experiment comparisons imprecise. Thus, a
mechanism is required to accurately align the data collected from each
of the synchrony/time-series experiments. The model we describe addresses
both of these problems.

Most of the numerous models designed to measure cell population
dynamics in synchrony/time-series experiments fall into two related
classes: population balance (PB) and branching process (BP) models.
PB models are usually formulated as partial-integro-differential
equations and are often very difficult to work with except under
special conditions~[Liou, Srienc and Fredrickson (\citeyear{liou1997spb}), Sidoli, Mantalaris and Asprey (\citeyear{sidoli2004})]. BP models are
stochastic models for population dynamics that have been used to study
both the asymptotic~[Alexandersson (\citeyear{alexandersson2001esb})] and short term
behaviors~[Larsson et~al. (\citeyear{larsson2008evs}), Orlando et~al. (\citeyear{CLOCCS})] of populations; certain BP
models have PB analogues~[Arino and Kimmel (\citeyear{arinokimmel93})]. Several models that
do not explicitly account for reproduction, and hence are neither PB
or BP models, have also been used to model data from asynchrony
experiments~[Bar-Joseph et~al. (\citeyear{barjoseph2004dcc}), Lu et~al. (\citeyear{luetal2004})].

The most critical distinction between models, however, is in the
sources of synchrony loss the model includes. Most describe synchrony
loss as the result of a single parameter, equivalent to a distribution
over division times
[Bar-Joseph et~al. (\citeyear{barjoseph2004dcc}), Chiorino et~al. (\citeyear{chiorino2001drc}), Larsson et~al. (\citeyear{larsson2008evs})]. In contrast,
the model we describe here (the CLOCCS model, in reference to its
ability to Characterize Loss of Cell Cycle Synchrony~[Orlando et~al. (\citeyear{CLOCCS})])
is the only model to account for variability in cell-division time,
initial asynchrony in the starting population and variability due to
asymmetric cell division~[Chiorino et~al. (\citeyear{chiorino2001drc})], all of which we will
show to be important. The CLOCCS model is based on a novel branching
process construction and can be written in closed form. Its
formulation is constructive, based on an accounting of unique cohorts
in the population at any given time. Hence, the model can attribute
one-time effects to specific subsets of the population, demonstrating
flexibility not available using the PB and BP models historically
applied to these populations. Further, the model's construction
allows full Bayesian inference without the use of approximations to
the likelihood. The Bayesian approach to inference has the additional
advantage that it sidesteps many of the difficulties encountered by
frequentist inference for BP models~[Guttorp (\citeyear{guttorp91})].

%
\begin{figure}[b]

\includegraphics{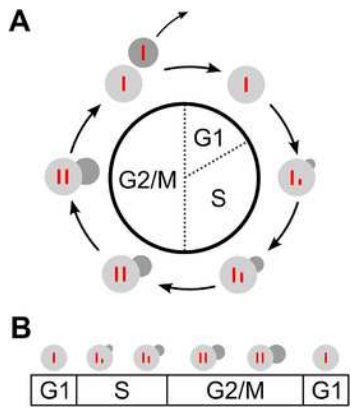}

\caption{Over the course of its life, the cell repeatedly traverses
the cell cycle, which is divided by landmark events associated with
asexual reproduction into the G1, S and G2/M phases. In the figure,
this corresponds to the cell in light gray traveling around the
circle in \textup{A} or from left to right in \textup{B}. At each
completion of G2/M it spawns a daughter cell. This process begins
with development of a bud (dark gray) and the start of DNA
replication (denoted by the appearance of a second red bar) and is
completed when the daughter cell (dark gray) separates from the
mother cell at the end of G2/M with a full complement of DNA.}\label{fig:basicCycleFig}
\end{figure}

In this paper we present a model which can utilize two forms of data
that provide information regarding the cell cycle position of \textit{Saccharomyces cerevisiae}, baker's yeast: DNA content data and budding index
data. An overview of the yeast cell cycle and these data types can be
found in Section~\ref{sec:yeast}. While applied here to yeast, the
ability to fit DNA content data, described in Section~\ref{sec:model},
is a
critical advance that allows the CLOCCS model to be applied to an
array of more complex organisms that do not undergo the kinds of
morphological changes that yeast do (e.g., budding) during the cell
division cycle. In Section~\ref{sec:analysis} we apply the model to
fit budding index and DNA content data from a synchrony/time-series
experiment in yeast. Using these data, we compare the model to a
collection of nested alternative parameterizations with subsets of the
novel asynchrony sources removed. We conclude with a discussion of
the model and of the results of this analysis in
Section~\ref{sec:disc}.

\section{Yeast cell cycle}\label{sec:yeast}
One organism commonly studied using synchrony/time-series experiments
is the common baker's yeast, \textit{S.~cerevisiae}, because many
features of its cell
cycle are well characterized. Figure~\ref{fig:basicCycleFig}A depicts
the landmark events that can be used to determine the cell cycle state
of individual cells [Gordon and Elliott (\citeyear{gordon1977fsc}), Hartwell (\citeyear{hartwell1974scc})]. The
first, bud emergence, is a distinct morphological landmark easily
detected by simple light microscopy. It first appears near the time
that a cell transitions from G1 into S phase. Cells become unbudded
after the completion of mitosis (M) when the cell and its bud
separate. We refer to the progenitor cell as the ``mother'' and what
had been the bud as the ``daughter.'' In~\textit{S.~cerevisiae}, this
division is
often asymmetric: the mother cell is often larger and progresses more
quickly through the cell cycle than the
daughter~[Hartwell and Unger (\citeyear{hartwell1977uds})]. Cell cycle position can also be
determined by measuring genomic DNA content of the cell, which
increases as cells progress through the S phase of the cell
cycle~[Haase and Reed (\citeyear{Haase02})]. Haploid yeast cells begin the cell cycle with
one copy of genomic DNA (red bar in Figure~\ref{fig:basicCycleFig}).
During the S phase, DNA is replicated such that, at the completion of
the S
phase, the cell has two copies of genomic DNA.

Counts of budded cells and cell-level DNA content are typically
measured in independent samples, drawn at regular time points after
the population's release from synchrony. The resulting time series of
budded cell counts is referred to as a budding index. DNA content is
measured by flow cytometry. Budding index and DNA content data can be
used to fit accurate models of the underlying cell cycle position
distributions.

\section{Model}\label{sec:model}
The model we describe is comprised of two components: an underlying
model for the population dynamics of the cells in a synchrony/time-series experiment, and independent sampling models for the
budding index and DNA content measurements made on samples drawn from
the population. We refer to the population dynamics model component
as CLOCCS. CLOCCS is a branching process model for position, $P_t$,
of a randomly sampled cell in a linearized version of the cell cycle
(Figure~\ref{fig:basicCycleFig}B)---which we refer to as a
cell cycle lifeline---given the experimental time, $t$, at which the
cell was sampled. The sampling models for the budding index and
DNA content measurements are conditioned on the distribution of lifeline
position and time. In what follows, we describe the model's
components in greater detail.

\subsection{Model for position given time}~\label{subsec:cloccs} The
CLOCCS model specifies the distribution of cell positions over an
abstract cell cycle lifeline as a function of time. We define
$\lambda$ to be the amount of time, in minutes, required by a typical
mother cell to undergo one full cell cycle. We divide the lifeline
into $\lambda$ units, thus the average cell will move one lifeline
unit per minute. The advantage of using a lifeline
characterization is that it allows for introduction of one-time
effects, such as the recovery period following release from synchrony
or the delay in cell cycle progression of new daughter cells.

We model position as having three independent sources of variability:
the velocity with which the cell traverses the cell cycle, the time it
spends recovering from the synchronization procedure, and the
additional time spent by a daughter cell as it traverses its first
cell cycle~[Hartwell and Unger (\citeyear{hartwell1977uds})]. It is well known that cells in
synchrony experiments progress through the cell cycle with varying
speeds. We assume that each cell moves at a constant velocity along
the lifeline, and that this velocity is random, following a normal
distribution. While this is technically inappropriate as velocities
must be positive, in practice it is reasonable: fitted distributions
give almost no mass to the negative half line. We measure velocity,
$V$, in lifeline units per minute; by definition, the mean cell
velocity is 1.0. The velocity distribution's variance, $\sigma^2_v$,
is unknown.

When released from synchrony, cells spend more time in their first G1
phase than they spend in G1 during subsequent cell cycles. The added
time reflects a period of recovery from the synchronization process,
whose length varies from cell to cell. We term this recovery period
Gr as if it were a distinct cell cycle phase. We model this effect as
a random offset, $P_0$, in the starting position on the lifeline.
While this offset should be strictly positive, we let $P_0$ be
distributed N($\mu_0,\sigma^2_0$) for convenience. Later, we
comment further on this choice. Daughter cells tend to be smaller and
require additional time in G1 before they begin to divide. We term
this daughter-specific period of growth Gd and model it by introducing
a fixed offset, $\delta$, to the cell's lifeline position.

%
\begin{figure}

\includegraphics{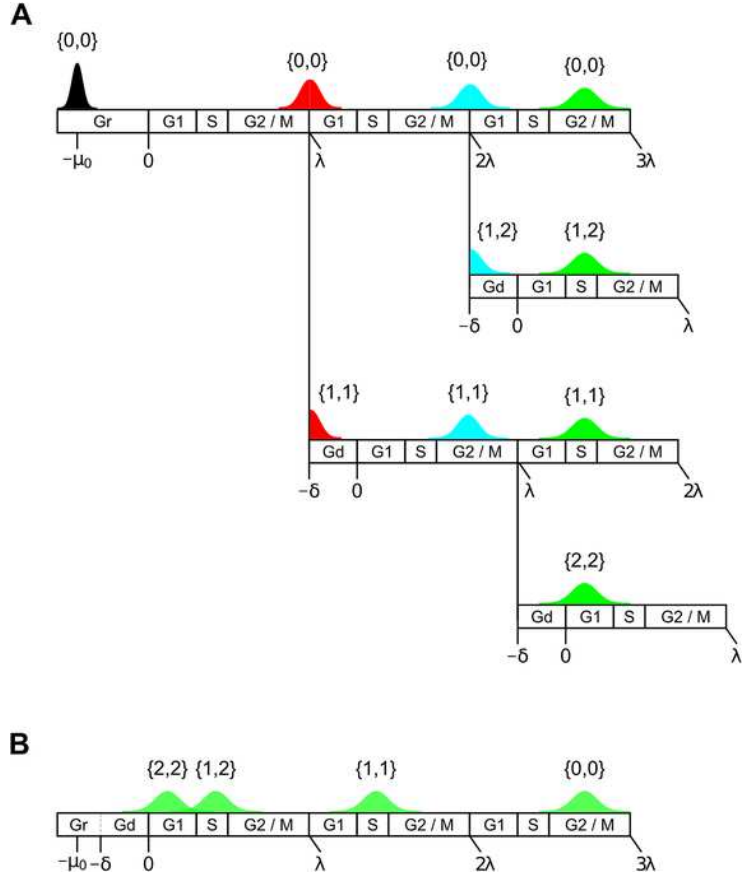}

\caption{Graphical representation of the branching dynamics of the cell
population, \textup{A}, and a snapshot in time plotted on a common
lifeline, \textup{B}. \textup{A}: The position distributions of
the cohorts, indexed by $\{g,r\}$, in the population at four points
in time, each color-coded. Black: at release from synchrony, there
is a single cohort, $\{0,0\}$. Red: as it enters its second cell
cycle, $\{0,0\}$ spawns a daughter cohort, $\{1,1\}$, located on its
own lifeline in Gd. Blue: at cohort $\{0,0\}$'s second reproductive
instance, it gives rise to another first generation cohort,
$\{1,2\}$; meanwhile, most cells in cohort $\{1,1\}$ are progressing
through G2/M. Green: the population is comprised of four distinct
cohorts; \textup{B}: a plot of the population at this time point on a
common lifeline.}\label{fig:BranchingFig}
\end{figure}

With each wave of division, the population expands in size. If cells
in the culture remained synchronous, the population would branch and
double in size every $\lambda$ minutes after an initial delay of
$\mu_0$ minutes. Because they do not, the dynamics of this expansion
is more complex: at any point in time, the population may represent a
number of distinct cohorts, each defined by its lineage. Cohorts are
determined by $g$, their ``generation''---the number of daughter stages
in their lineage---and $r$, their ``reproductive instance''---the wave
of division that gave rise to the cohort.
Figure~\ref{fig:BranchingFig}A depicts the branching dynamics of this
process and a snapshot in time projected onto a common lifeline
(Figure~\ref{fig:BranchingFig}B). In A, four distinct time periods
are color coded with each cohort distribution labeled with its
$\{g,r\}$ index. At time zero there is a single cohort, $\{0,0\}$,
depicted in black, whose position distribution is located in Gr and
centered at $-\mu_0$. As time passes (red), this cohort enters its
second cell cycle and spawns a daughter cohort, labeled $\{1,1\}$,
which begins on its own lifeline in Gd. Later (blue), cohort
$\{0,0\}$ gives rise at its second reproductive instance to another
first generation cohort, $\{1,2\}$. At the same time, cohort
$\{1,1\}$ cells are progressing through G2/M. At the last depicted
time point (green), the population is comprised of four distinct
cohorts, representing three generations of cells arising at three
distinct reproductive instances. Figure~\ref{fig:BranchingFig}B is a
plot of the population at this time point on the common lifeline. The
CLOCCS model is a distribution over position along this common
lifeline as a function of time.

In what follows we use a description of the behavior of individual
cells as a device for deriving population level cohort position
distributions. Each such distribution is normal with parameters that
depend on the starting position and velocity distributions, time $t$
and the cohort's indices $g$ and $r$. Since cells in the $\{0,0\}$
cohort are unaffected by the daughter specific delay, $\delta$, their
positions, $P_t$, at time $t$ are determined only by their starting
positions, $P_0$, and their velocity, $V$. For these cells, $P_t = P_0
+ Vt$. Hence, $\mathrm{p}(P_t|\Theta,R=0, G=0, t)$ is normal with mean
$-\mu_0+t$ and standard deviation $\sqrt{\sigma_0^2+t^2\cdot\sigma
_v^2}$. In contrast, cells
in cohorts at generations greater than zero have their position
distributions truncated at the beginning of Gd, $-\delta$ on the
lifeline, and are set back by $g$ daughter offsets of length $\delta$
and $r$ cell cycle offsets of length $\lambda$. The remaining
contributions to such a cell's position are the velocity by time
contributions of each of its ancestors and the initial position of its
ancestor in cohort $\{0,0\}$. For simplicity, we assume that daughter
cells inherit their mother cell's velocity. With this, the velocity
by time contribution to position simplifies to $Vt$, where $V$ is the
common velocity and $t$ is total time since population release. For
these cells, $P_t = P_0 + Vt - g\delta- r\lambda$, hence,
position, $\mathrm{p}(P_t|\Theta, G\ge0, R\ge G, t)$, is normal with mean
$-\mu_0+ t - g \delta- r \lambda$ and standard deviation
$\sqrt{\sigma_0^2+t^2\cdot\sigma_v^2}$ truncated so that $P_t \ge-
\delta$.

Thus, we write the model for position, $P_t$, given time, $t$, in
closed form by enumerating the population's cohorts using the latent
variables $r$ and $g$. In particular,
%
%
\begin{equation} \label{eqn:cloccs}
\mathrm{p}(P_t|\Theta,t) = \sum_{\mathcaligr{C}} \mathrm{p}(P_t|\Theta,r,g,t)
\mathrm{p}(g,r|\Theta,t),
\end{equation}
where $\Theta= (\mu_0, \sigma^2_0, \sigma^2_v, \lambda, \delta)$
and where the sum is over possible cohorts, $\mathcaligr{C} = \{ \{g,r\}
{}\dvtx{} (g=0 \wedge r=0) \vee(0 < g \le r \le R) \}$. While the number of
cohorts represented in the population could theoretically be large, in
practice, their number is limited by the number of cell cycles that
cohort $\{0,0\}$ is able to undergo during the experimental period.
In most cases, synchrony experiments are terminated after 2 or 3
cycles, so choosing $R=4$, 5 or 6 is usually sufficient. For
notational clarity, we use $C$ to represent the sufficient number of
cell cycles examined.

The marginal probability of drawing a representative of cohort $\{g,
r\}$ from the population at time $t$ is $\mathrm{p}(g,r|\Theta,t)$. For
example, in the scenario depicted in Figure~\ref{fig:BranchingFig}B,
$\mathrm{p}(1,1|\Theta,t)$ is the ratio of the mass under the cohort
$\{1,1\}$ density to the total mass under all of the cohort densities
present on the lifeline. The mass under the cohort $\{1,1\}$ density
is the probability that a randomly drawn member of the $\{0,0\}$
cohort has completed its first cell cycle and contributed a daughter
cell to cohort $\{1,1\}$. This probability is equal to $ (1 -
\Phi( \frac{\mu_0- t + \lambda
}{\sqrt{\sigma_0^2+t^2\cdot\sigma_v^2}}))$. The mass
under the cohort $\{2,2\}$
density is the probability that a randomly drawn member of the
$\{1,1\}$ cohort has finished its first cell cycle; this, in turn, is
the probability that a randomly chosen member of the $\{0,0\}$ cohort
has traveled $\delta$ units into its third cell cycle. The $\delta$
appears because the $\{1,1\}$ cohort's progress through its first cell
cycle is $\delta$ units longer than the $\{0,0\}$ cohort's progress
through its second cell cycle. In this way, the relative contribution
of any cohort in the population can be determined by calculating
the probability that the position of a randomly drawn member of the
$\{0,0\}$ cohort is past a threshold position that is a function of
$g$ and~$r$. Let $M_\Theta(g,r,t)$ denote the mass under cohort
$\{g, r\}$'s position distribution at time $t$,
\begin{eqnarray*}
&&M_\Theta(g,r,t)
\\
&&\quad=\cases{
1, &\quad $g=0, r=0$, \cr\vspace*{2pt}
\biggl(1-\Phi\biggl(\displaystyle\frac{\mu_0-t+r\cdot\lambda+(g-1)\cdot\delta}{\sqrt{\sigma_0^2+t^2\cdot\sigma_v^2}}\biggr)\biggr)\cdot{{r-1}\choose{g-1}},
&\quad  $g\geq1,r\geq g$,\cr\vspace*{2pt}
0, &\quad else,
}
\end{eqnarray*}
where $\Phi(\cdot)$ denotes the standard normal CDF. The combinatoric term
arises from that fact that, for $r\geq g \geq1$, multiple lineages
may contribute members to a given cohort. For example, cohort
$\{1,1\}$ will contribute
to cohort $\{2,3\}$ as its members pass the
point $2\lambda$ on its lifeline (rightmost point on the third branch
from top in Figure~\ref{fig:BranchingFig}A), while cohort $\{1,2\}$
will contribute to the same cohort, $\{2,3\}$, as its members pass the
point $\lambda$ on its lifeline (rightmost point on the second branch
from top in Figure~\ref{fig:BranchingFig}A). Finally, let $Q_\Theta
(t)$ denote the mass under all cohort distributions in the population
at that time,
\[
Q_\Theta(t) = \sum_{\mathcaligr{C}} M_\Theta(g,r,t).
\]
In general, $\mathrm{p}(g,r|\Theta,t) = M_\Theta(g,r,t)/Q_\Theta(t)$.

%
\begin{figure}[b]

\includegraphics{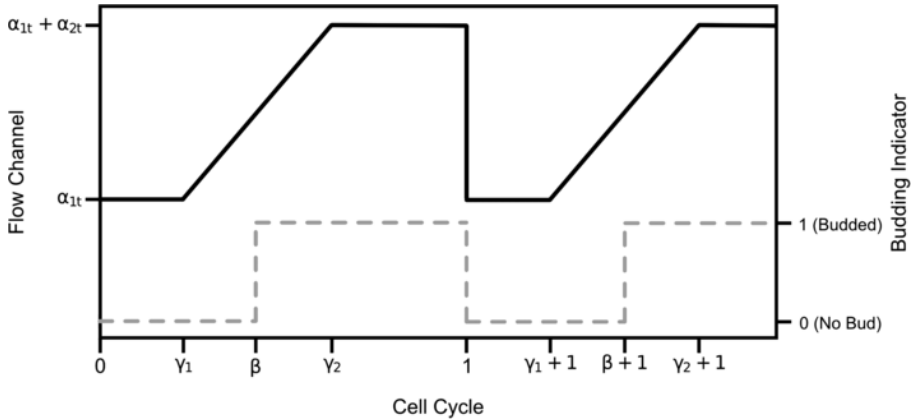}

\caption{Plot of expected flow cytometry channel for a cell given its
lifeline position in units of $\lambda$ (black curve, left vertical
axis). An indicator function for the cell's budding status is also
plotted (grey dashed curve, right vertical
axis).}\label{fig:FlowFig}
\end{figure}

\subsection{Sampling models}~\label{subsec:budflow} To utilize the
CLOCCS model, it is necessary to relate distributions over the
artificial cell cycle lifeline to observable cell features. In the
next two sections we present two sampling models which allow CLOCCS to
utilize commonly collected landmark data, namely budding index and DNA content
data. While time series of budding index and DNA content data are each
sufficient to estimate the CLOCCS parameters, $\Theta$, they provide
complementary information on the cell cycle timing of distinct
landmark events. Timing of these events is of independent interest,
and estimates of the same may improve the utility of the model as a
tool for deconvolution of transcription data and other types of
downstream analysis.

\subsection{Sampling model for budding index
data}~\label{subsec:budlike} Presence or absence of a bud is an
easily measured landmark tied to a cell's progression through the cell
cycle (see Figure~\ref{fig:basicCycleFig}). Buds emerge and become
detectable near the transition between G1 and S phases, at a fraction
$\beta$ of the way through the normal cell cycle and split off as
daughter cells at cell cycle completion (Figure~\ref{fig:FlowFig},
dashed line).

Assume that budding index samples are drawn at $T$ time points,
$t_i, i=1,\ldots,T$, and that $n_i$ cells are counted at time $t_i$.
Let $b_{ji}=1$ if the $j$th cell at time $t_i$ is budded and
$b_{ji}=0$ otherwise. The event that $b_{ji}=1$ implies that the
position of the $j$th cell at time $t_i$, $P_{ji}$, falls into the
lifeline interval $((c+\beta)\lambda,(c+1)\lambda]$ for
some cell cycle $c \ge0$; the probability of this is dictated by the
CLOCCS model.

Following the development of Section~\ref{subsec:cloccs}, we calculate
$\mathrm{p}(b_{ji}=1|\beta,\Theta,t_i)$ by introducing cohorts and
marginalizing over them. In particular, let
\[
\mathrm{p}(b_{ji}=1|\beta,\Theta,t_i) = \sum_{\mathcaligr{C}}
\mathrm{p}(b_{ji}=1|\beta,\Theta,g,r,t_i) \mathrm{p}(g,r|\Theta,t_i),
\]
where $\mathrm{p}(b_{ji}=1|\beta,\Theta,g,r,t_i)$ is the probability
that a
cell randomly sampled from cohort $\{g,r\}$ is budded at time $t_i$.
For the progenitor cohort, $\{0,0\}$,
\begin{eqnarray*}
&&\mathrm{p}(b_{ji}=1|\beta,\Theta,g,r,t_i)
\\
&&\quad= \sum_{c=0}^{C}\biggl[\Phi\biggl(\frac{\lambda\cdot(c+1)-(-\mu_0+t_i)}{{\sqrt{\sigma_0^2+t_i^2\cdot\sigma_v^2}}}\biggr)
- \Phi\biggl(\frac{\lambda \cdot(c + \beta)-(-\mu_0+ t_i)}{{\sqrt{\sigma_0^2+t_i^2\cdot\sigma_v^2 }}}\biggr)\biggr],
\end{eqnarray*}
while, for subsequent cohorts, $0 < g \le r$,
\begin{eqnarray*}
&&\mathrm{p}(b_{ji}=1|\beta,\Theta,g,r,t_i) \\
&&\quad= \sum_{c=0}^{C}\biggl[\biggl(\Phi\biggl(\frac{\lambda\cdot(c +1)
-({-\mu_0+t_i-r\cdot\lambda-g\cdot\delta})}{{\sqrt{\sigma_0^2+
t_i^2\cdot\sigma_v^2}}}\biggr)
\\
&&\qquad\hspace*{26pt}{}- \Phi\biggl(\frac{\lambda\cdot(c + \beta)
-({-\mu_0+t_i - r\cdot\lambda-g\cdot\delta})}{\sqrt {\sigma_0^2
+ t_i^2\cdot\sigma_v^2}}\biggr)\biggr)
\\
&&\hspace*{38pt}\qquad{}\Big/
\biggl(1-\Phi\biggl(\frac{-\delta-({-\mu_0+
t_i-r\cdot\lambda-g\cdot\delta})}{{\sqrt{\sigma_0^2+ t_i^2\cdot\sigma_v^2}}}\biggr)\biggr)\biggr].
\end{eqnarray*}

We model bud presence as a Bernoulli random variable with success
probability $\mathrm{p}(b_{ji}=1|\beta,\Theta,t_i)$ and assume that samples
drawn at the various time periods are independent conditional on the
CLOCCS model.

\subsection{Sampling model for DNA content data}~\label{subsec:flowlike}
DNA content data measured by flow cytometry provides an ordinal measurement
of the DNA content of each cell in a sample: each cell appears in one
of 1024 ordered channels on the basis of its fluorescence, which is
proportional to its DNA content~[Pierrez and Ronot (\citeyear{pierrezronot92})]. In practice,
channel number is often log$_2$ transformed and treated as a
continuous measurement.

Adapting the CLOCCS model to DNA content data requires that we annotate the
lifeline with the positions, measured as fraction of cell cycle
length, at which S phase begins and ends. We denote these locations
$\gamma_1$ and $\gamma_2$, respectively. As the population loses
synchrony, the distribution of cells over channels will typically be
bimodal, with one mode corresponding to cells in G1 (centered at
$\alpha_1$), and the another corresponding to G2/M (centered at
$\alpha_1+\alpha_2$). Cells transiting the S phase will fall between these
points in expectation. Further, we assume that DNA content increases
linearly over the course of the S phase. In particular, the expected DNA
content of a cell is
%
\begin{equation} \label{eqn:dnaCurveFunc}
\sum_{c=0}^C
\cases{
\alpha_{1t},\qquad\hspace*{53pt} c\lambda\leq P_t < (c + \gamma_{1})\lambda,\qquad\hspace*{31pt} \mbox{G1},\cr
\omega_{1t} P_t + \omega_{0t}(c),\qquad (c + \gamma_{1})\lambda
\leq P_t < (c + \gamma_{2})\lambda,\qquad \mbox{S-Phase},\cr
\alpha_{2t}+\alpha_{1t},\qquad\hspace*{25pt} (c + \gamma_{2})\lambda\leq P_t <(c+1)\lambda,\qquad\hspace*{6pt} \mbox{G2/M},
}
\end{equation}
where $ \omega_{1t}=
\frac{\alpha_{2t}}{\lambda(\gamma_{2}-\gamma_{1})}$ and $
\omega_{0t}(c)= \frac{\alpha_{1t}(\gamma_{2}- \gamma_{1}) -
\alpha_{2t}(\gamma_{1}+ c)}{\gamma_{2}- \gamma_{1}}$.
The black line in Figure~\ref{fig:FlowFig} is a plot of this curve.

Measurement of DNA content by flow cytometry is imprecise. Machine
noise, variation in the cell's orientation to the laser beam and
variation in the performance of the fluorescent stain each contribute
to measurement error~[Pierrez and Ronot (\citeyear{pierrezronot92})]. Hence, a flow
cytometry measurement made on a sample of cells drawn at a particular
time point will be a sample from the convolution of a noise
distribution and the CLOCCS position distribution. In particular,
%
\begin{equation} \label{eqn:flowconvolution}
\mathrm{p}(f_{ji}|\Psi, \Theta,t) = \int_{-\infty}^{\infty}
\mathrm{p}(f_{ji}|P_t,\Psi, \Theta, t)
\mathrm{p}(P_t|\Theta,t)\, dP_t,
\end{equation}
where $f_{ji}$ denotes the log fluorescence intensity of cell $j$ at
time $t_i$ and where $\Psi$ denotes the vector of parameters in the
model for $f_{ji}$ not in $\Theta$. From above it follows that
$\mathrm{p}(f_{ji}|P_t,\Psi,\Theta,t)$ can be modeled as a normal
with mean
given in equation~(\ref{eqn:dnaCurveFunc}) and variance $\tau_t^2$. The
log normal distribution is a common choice in this setting
[Gray and Dean (\citeyear{gray1980daa})]. Additionally, the noise characteristics of the
flow cytometer typically vary from one sample to the next, causing the
locations of the G1 and G2/M modes, as well as the level of machine
noise ($\tau$) to vary. Hence, we allow the parameters of the DNA content
sampling distribution, $\mathrm{p}(f_{ji} |P_t,\Psi, \Theta, t )$,
to vary
across time periods.

Note that equation~(\ref{eqn:flowconvolution}) can be written as
\[
\mathrm{p}(f_{ji}|\Psi, \Theta,t) = \sum_{\mathcaligr{C}} I_{gr}(f_{ji})
\mathrm{p}(g,r|\Theta,t),
\]
where
\[
I_{gr}(f_{ji}) = \int_{-\infty}^{\infty} \mathrm{p}(f_{ji}|P_t,\Psi, \Theta, t)
\mathrm{p}(P_t|\Theta,r,g,t)\, dP_t
\]
is a convolution of two normals, one of which is truncated.

Let $l_{gr}$ denote the left limit to the support of cohort
$\{g,r\}$'s position distribution, where $l_{gr}=-\infty$ if $g=r=0$
and $l_{gr}=-\delta$ otherwise. Further, let $G_{grt}(x)$ denote the
normal cumulative distribution function with mean $-\mu_0+t-r\cdot
\lambda-g\cdot\delta$ and
variance $\sigma_0^2+t^2\cdot\sigma_v^2$ evaluated at $x$ and let
$S_{cgrt}(x)$ denote
the normal cumulative distribution function with mean
\begin{eqnarray*}
&&\biggl((\sigma_0^2+t^2\cdot\sigma_v^2)
\biggl( \frac{f_{ji}}{\omega_{1t}} -
\frac{\omega_{0t}(c)}{\omega_{1t}}  \biggr)
+  \biggl(\frac{\tau^2}{\omega_{1t}^2}  \biggr)
(-\mu_0+t-r\cdot\lambda-g\cdot\delta)\biggr)
\\
&&\qquad{}\Big/\biggl(\sigma_0^2+t^2\cdot\sigma_v^2+
\frac{\tau^2}{\omega_{1t}^2}\biggr)
\end{eqnarray*}
and variance
\[
\biggl((\sigma_0^2+t^2\cdot\sigma_v^2) \frac{\tau^2}{\omega
_{1t}^2}\biggr)\Big/\biggl(\sigma_0^2+t^2\cdot\sigma_v^2+
\frac{\tau^2}{\omega_{1t}^2}\biggr)
\]
evaluated at $x$. It can be shown that $I_{gr}(f_{ji}) =
I^{*}_{gr}(f_{ji})/(1-G_{grt}(l_{gr})), $ where
\begin{eqnarray*}
&&I^{*}_{gr}(f_{ji})  \\
&&\quad= \frac{1}{\tau} \phi \biggl(\frac{f_{ji}-\alpha_{1t}}{\tau}  \biggr)
\\
&&\qquad{}\times\Biggl[\bigl(G_{grt}(\gamma_1\lambda)-G_{grt}(l_{gr})\bigr)+\sum_{c=1}^{C}\bigl(G_{grt}\bigl((c+\gamma_1)\lambda\bigr)-G_{grt}(c\lambda)\bigr)\Biggr] \\
&&\qquad{} + \frac{1}{\tau}\phi
\biggl(\frac{f_{ji}-\alpha_{1t}-\alpha_{2t}}{\tau}  \biggr)
\sum_{c=0}^{C} \bigl(G_{grt}\bigl((c+1)\lambda\bigr)-G_{grt}\bigl((c +
\gamma_2)\lambda\bigr)\bigr)
\\
&&\qquad{}+ \sum_{c=0}^{C}
 \Biggl(\phi \Biggl( \biggl(\frac{f_{ji}}{\omega_{1t}}
-\frac{\omega_{0t}(c)}{\omega_{1t}} - (-\mu_0+t-r\cdot\lambda
-g\cdot\delta)\biggr)
\\
&&\qquad\hspace*{139pt}{}\bigg/\sqrt{\sigma_0^2+t^2\cdot\sigma_v^2+
\frac{\tau^2}{\omega_{1t}^2}}  \Biggr)
\\
&&\hspace*{132pt}\qquad{}
\Big/{ \sqrt{ \omega_{1t}
^2 (\sigma_0^2+t^2\cdot\sigma_v^2) +\tau^2}}\Biggr)
\\
&&\qquad\hspace*{27pt}{}\times \bigl(S_{cgrt}\bigl((c + \gamma_2)\lambda\bigr) - S_{cgrt}\bigl((c + \gamma_1)\lambda\bigr) \bigr)
\end{eqnarray*}
and where $\phi(\cdot)$ is the standard normal density function.
In the equation above, the first line of the right-hand side corresponds
to cells in G1, the second to cells in G2 or M, and the third to cells
in S.


We assume that cell-level DNA content measurements are conditionally
independent within and between samples drawn at the various time
periods conditional on the CLOCCS model, $\Psi$ and the sampling
times. DNA content and budding index measurements are made on separate
samples drawn from a population's culture, sometimes at the same
points in time, sometimes not. Because they are distinct samples, we
model the DNA content and budding index data as conditionally
independent given the CLOCCS parameters $\Theta$, the budding
parameter $\beta$, the DNA content parameters $\Psi$ and sampling times.


\subsection{Prior distribution}~\label{subsec:prior} What follows is a
description of, and justification for, the prior choices used in our
analysis. Columns 2 and 3 of Table~\ref{tab:estimates} tabulate prior
expected values and 95\% equal-tailed intervals for each parameter as
implied by these choices.

Lord and Wheals (\citeyear{lordwheals83}) estimate \textit{S.~cerevisiae} cell cycle
length in culture at
30 degrees Centigrade---the temperature employed by our lab---to be
78.2 minutes with a standard deviation of 9.1 minutes. To allow for
differences in experimental protocol, we place a normal, mean 78.2,
standard deviation 18.2 prior on cell cycle length, $\lambda$. In
\textit{S.~cerevisiae}, duration of the S phase, $(\gamma_2 - \gamma_1)\lambda$, is about
one quarter of the cell cycle; it begins a short time before buds can
be visually detected and continues until mother and daughter cells
separate~[Vanoni, Vai and Frascotti (\citeyear{vanoni1984ety})]. Based on an analysis of 30 DNA content
measurements made on an asynchronous population conducted using the
same protocol as used in the synchrony experiment described in the
next section, we estimate that $\gamma_1$ is approximately 0.1 and
that $\beta$ is approximately 0.12. Hence, we expect $\gamma_1 <\beta< \gamma_2$.
With this in mind, we let $\gamma_1 \sim\operatorname{Beta}
(2,18)$, $\beta\sim\operatorname{Beta}(2.4,17.6)$ and $ \gamma_2 \sim\operatorname{Beta}(7,13)$,
constrained as above. Bar-Joseph et~al. (\citeyear{barjoseph2004dcc})
estimates the standard deviation of the velocity distribution in
\textit{S.~cerevisiae} to be 0.09 and observed a range of values 0.07
to 0.11 across 3 experiments. For this reason, we place an independent
$\operatorname{inverse}$-$\operatorname{gamma}(12,1)$ prior distribution on $\sigma_v$.

Aspects of experimental protocol, most notably the method used to
synchronize the population, have a strong influence on the parameters
of the starting position distribution and on duration of the
daughter-specific offset, $\delta$. Centrifugal elutriation, the
method used in the experiment we describe in the next section, selects
for small unbudded cells, while other methods, such as $\alpha$-factor
arrest, do not. Because of their size, elutriated cells tend to spend
more time in Gr and their daughters spend more time in Gd than their
counterparts in $\alpha$-factor experiments [Hartwell and Unger (\citeyear{hartwell1977uds})].
We have chosen to specify our prior distributions on these parameters
to accommodate---not condition on---this source of protocol dependent
uncertainty. In particular, we place an inverse-gamma distribution
with shape parameter 2 and mean 78.2$/$3 on $\sigma_0$ and the minimally
informative exponential, mean 78.2 prior distribution on $\mu_0$. The
former reflects our belief that almost all cells will be in Gr at
release; the latter places highest prior likelihood on a short Gr, as
is expected in an $\alpha$-factor experiment, but allows for the
longer Gr that is expected in elutriation experiments. Similar
reasoning was behind our choice of an exponential mean 55 prior
distribution on $\delta$: in $\alpha$-factor experiments, $\delta$ can
be very brief, while in elutriation experiments it can exceed 40\% of
the length of a typical cell
cycle~[Hartwell and Unger (\citeyear{hartwell1977uds}), Lord and Wheals (\citeyear{lordwheals83})].

%
\begin{sidewaystable}
\tablewidth=\textwidth
\caption{Prior (columns \textup{1} and \textup{2}) and marginal posterior summaries given
both the DNA content and budding index data (columns \textup{3} and \textup{4}), given the
DNA content data only (columns \textup{5} and \textup{6}) and given the budding index data
only (columns \textup{7} and \textup{8})}\label{tab:estimates}
\begin{tabular*}{\textwidth}{@{\extracolsep{\fill}}ld{2.3} cd{3.3} cd{3.3} cd{3.3}cc@{}}
\multicolumn{9}{@{}c}{\hrulefill}\\
& \multicolumn{2}{c}{\textbf{Prior}}
& \multicolumn{2}{c}{\textbf{Flow and budding}}
& \multicolumn{2}{c}{\textbf{Flow only}}
& \multicolumn{2}{c}{\textbf{Budding only}} &
\\[-6pt]
& \multicolumn{2}{c}{\hrulefill}
& \multicolumn{2}{c}{\hrulefill}
& \multicolumn{2}{c}{\hrulefill}
& \multicolumn{2}{c}{\hrulefill} &
\\
& \multicolumn{1}{c}{\textbf{Mean}}
& \textbf{95\% interval}
& \multicolumn{1}{c}{\textbf{Mean}}
& \textbf{95\% interval}
& \multicolumn{1}{c}{\textbf{Mean}}
& \textbf{95\% interval}
& \multicolumn{1}{c}{\textbf{Mean}}
& \textbf{95\% interval} & \\[-6pt]
\multicolumn{9}{@{}c}{\hrulefill}
\\
$\mu_0$&78.200&(22.497, 288.470)&-94.387&($-$94.778, $-$94.000)&-94.279&($-$94.674, $-$93.878)&-95.967&($-$98.770, $-$92.915)&$\rceil$ \\
$\delta$&55.000&(15.823, 202.888)&44.318&(43.842, 44.790)&44.326&(43.845, 44.810)&41.748&(36.623, 46.424)&\multirow{3}{7pt}{$\Theta$}\\
$\sigma_0$&26.066&(4.679, 107.150)&17.961&(17.765, 18.150)&18.015&(17.819, 18.207)&15.328&(13.992, 16.555)&\\
$\sigma_v$&0.091&(0.051, 0.162)&0.025&(0.020, 0.029)&0.025&(0.020, 0.029)&0.058&(0.040, 0.079)&\\
$\lambda$&78.200&(65.924, 113.871)&79.487&(78.974, 79.995)&79.647&(79.133, 80.169)&76.785&(72.536, 81.354)&$\rfloor$
\\[3pt]
$\beta$&0.144&(0.045, 0.289)&0.153&(0.141, 0.165)& &&0.142&(0.114, 0.169)&
\\[3pt]
$\gamma_1$&0.068&(0.010, 0.168)&0.049&(0.046, 0.053)&0.050&(0.046, 0.054)&&&$\rceil$\\
$\gamma_2$&0.358&(0.181, 0.568)&0.349&(0.345, 0.352)&0.349&(0.345, 0.353)&&&\multirow{6}{8pt}{$\Psi$}\\
$\mu_{\alpha1}$&7.576&(6.477, 8.675)&8.237&(8.170, 8.304)&8.237&(8.170, 8.304)&&&\\
$\sigma^2_{\alpha1}$&0.065&(0.012, 0.269)&0.038&(0.023, 0.062)&0.038&(0.023, 0.062)&&&\\
$\mu_{\alpha2}$&0.818&(0.411, 1.224)&1.038&(0.841, 1.245)&1.035&(0.840, 1.237)&&&\\
$\sigma^2_{\alpha2}$&0.009&(0.002, 0.037)&0.334&(0.174, 0.598)&0.326&(0.171, 0.583)&&&\\
$\mu_{\tau}$&-1.906&($-$3.467, $-$0.346)&-2.094&($-$2.154, $-$2.033)&-2.094&($-$2.154, $-$2.033)&&&\\
$\sigma^2_{\tau}$&0.132&(0.024, 0.543)&0.031&(0.019, 0.050)&0.031&(0.019, 0.050)&&&$\rfloor$
\\[-6pt]
\multicolumn{9}{@{}c}{\hrulefill}
\end{tabular*}
\end{sidewaystable}

In the DNA content distributions, flow cytometer fluorescence noise, as
measured by $\tau_i$, and location of the G1 and G2/M modes, as
measured by $\alpha_{1i}$ and $\alpha_{2i}$ respectively, vary
randomly from assay to assay over time. We model this variability
hierarchically: first placing independent normal prior distributions
on $\log(\tau_{i})$, $\alpha_{1i}$, and $\alpha_{2i},  i=1,
\ldots, T$, followed by independent conjugate
normal-inverse-chi-square hyperprior distributions on the parameters
of the normal distributions. The latter are parametrized as
in~Gelman et~al. (\citeyear{gelmanetal95}). In particular,
\begin{eqnarray*}
\log(\tau_i) &\stackrel{\mathrm{i.i.d.}}{\sim}&\mathrm{N}(\mu_\tau,\sigma^2_\tau),\qquad\hspace*{44.5pt}
\alpha_{1i} \stackrel{\mathrm{i.i.d.}}{\sim}\mathrm{N}(\mu_{\alpha1},\sigma^2_{\alpha1}),
\\
\mu_\tau| \sigma^2_\tau&\sim&\mathrm{N}(\eta_\tau,\sigma^2_\tau/\kappa_\tau),\qquad\hspace*{12pt}
\mu_{\alpha1} | \sigma^2_{\alpha1} \sim\mathrm{N}(\eta_{\alpha1},\sigma^2_{\alpha1} / \kappa_{\alpha1} ),
\\
\sigma^2_\tau&\sim&\operatorname{Inv\mbox{-}\chi^2} (\nu_\tau,\gamma^2_\tau),\qquad\hspace*{26.5pt}
\sigma^2_{\alpha1} \sim\operatorname{Inv\mbox{-}\chi^2}
(\nu_{\alpha1},\gamma^2_{\alpha1}),\\
\alpha_{2i} &\stackrel{\mathrm{i.i.d.}}{\sim}&\mathrm{N}(\mu_{\alpha2},\sigma^2_{\alpha2}),\\
\mu_{\alpha2} | \sigma^2_{\alpha2} &\sim&\mathrm{N}(\eta_{\alpha2},\sigma^2_{\alpha2} / \kappa_{\alpha2}),\\
\sigma^2_{\alpha2} &\sim&\operatorname{Inv\mbox{-}\chi^2} (\nu_{\alpha2},\gamma^2_{\alpha2}),
\end{eqnarray*}
where $\operatorname{Inv\mbox{-}\chi^2} (\nu,\gamma^2)$ denotes the scaled inverse
$\chi^2$ distribution with $\nu$ degrees of freedom and scale
parameter $\gamma$. Given this specification, we define
\[
\Psi=
(\tau_1, \ldots, \tau_T, \alpha_{11}, \ldots, \alpha_{1T},
\alpha_{21}, \ldots, \alpha_{2T}, \mu_\tau, \sigma^2_\tau, \mu
_{\alpha
1}, \sigma^2_{\alpha1}, \mu_{\alpha2}, \sigma^2_{\alpha2} ).
\]

We chose the hyperparameters of the above hierarchical model on the basis
of an exploratory analysis of the same asynchronous DNA content data used
above. We set $\eta_{\alpha1} = 7.58$, $\eta_{\alpha2} = 0.82$ and
$\eta_\tau= -1.91$, the average of the observed estimates of
$\alpha_1$, $\alpha_2$ and $\tau$, respectively. We set each of the
prior sample size parameters, $\kappa_\tau$, $\kappa_{\alpha1}$ and
$\kappa_{\alpha2}$, and each of the prior degrees of freedom
parameters, $\nu_\tau$, $\nu_{\alpha1}$ and $\nu_{\alpha2}$, equal
to 2 to keep these margins of the prior distribution relatively
diffuse. Finally, we set $\gamma^2_\tau= 0.13$, $\gamma^2_{\alpha1}
= 0.065$ and $\gamma^2_{\alpha2} = 0.0089$---in each case 16 times
the observed variance in the asynchronous experiment.

\section{Analysis}\label{sec:analysis}
In what follows, we utilize the model to analyze budding index and
DNA content data from a cell cycle synchrony experiment in \textit{S.~cerevisiae} using
cells synchronized by centrifugal elutration and cultured at
30$^{\circ}$C. Details of the strain and growth conditions used can
be found in~Orlando et~al. (\citeyear{CLOCCS}). After synchronization, 32 samples were
collected at 8 minute intervals starting 30 minutes after release.
Two aliquots were taken from each sample, one for each type of
measurement. Budding index was measured by microscopically
assessing at least 200 cells for the presence of a bud and recording
the number of budded and unbudded cells observed. The relative DNA
content of 10,000 cells in each sample was measured by flow cytometry
as described previously~[Haase and Reed (\citeyear{Haase02})]. The observed fluorescence
values for each measured cell in each sample were $\log_2$ transformed
prior to analysis. The DNA content measurement of the 38 minute sample was
not available due to a technical problem encountered during
preparation of that sample.

We compare parameter estimates given both the budding index and
DNA content data, given the DNA content data alone and given the
budding index
data alone. In addition, using only the budding index data, we
estimate Bayes factors for the full CLOCCS model to submodels
obtained by systematically removing each novel source of asynchrony,
$\delta$, $\mu_0$ and $\sigma_0$ separately and in combination.

\subsection{Estimates given the experimental data}\label{subsec:fits} We use a random
walk\break
Metropolis~[Gilks, Richardson and Spiegelhalter (\citeyear{gilksetalintro1996b}), Metropolis et~al. (\citeyear{metroetal53})] algorithm for
each model fit. In each case, the algorithm was tuned to mix well and
the chain was given a lengthy burn-in period. Subsequent to this, we
ran the chain for 400,000 iterations and saved every fourth for
inference. Plots of sampled values appear stationary, and the Raftery
and Lewis diagnostic [Raftery and Lewis (\citeyear{raftlewi96})], implemented in the R
package CODA, indicates that the sample is sufficient to estimate the
$0.025$th quantile of any marginal posterior to within 0.01 with
probability 0.95. All coefficients and associated interval estimates
are based on summary statistics of marginal sample distributions. We
tested our implementation of the model and the Markov chain Monte
Carlo sampler by analyzing simulated data sets. Parameter estimates
derived from these analyses were consistent with their true values.

%


Table~\ref{tab:estimates} provides marginal summaries of the prior
(columns 1 and 2) and of the posterior distributions after fitting the
model to both the DNA content and budding index data (columns 3 and 4), to
the DNA content data only (columns 5 and 6) and to the budding index data
only (columns 7 and 8). Note that point and interval estimates of
common parameters derived using both the budding index and DNA content
data are very close to their counterparts fit only to the DNA content data.
This is not surprising given the information rich nature of the DNA content
data: at each time period approximately 10,000 cells are assayed for
DNA content, while only approximately 200 are assayed for presence of a
bud. On average, point estimates of the common parameters differ by
less than 1\% and the associated posterior interval estimates are only
about 2\% narrower when the budding index data is added. The
parameter $\beta$ can only be estimated with budding index data,
but it is estimated more accurately when DNA content data is included, owing
to the fact that it is constrained by $\gamma_1$ and $\gamma_2$.

%
\begin{figure}

\includegraphics{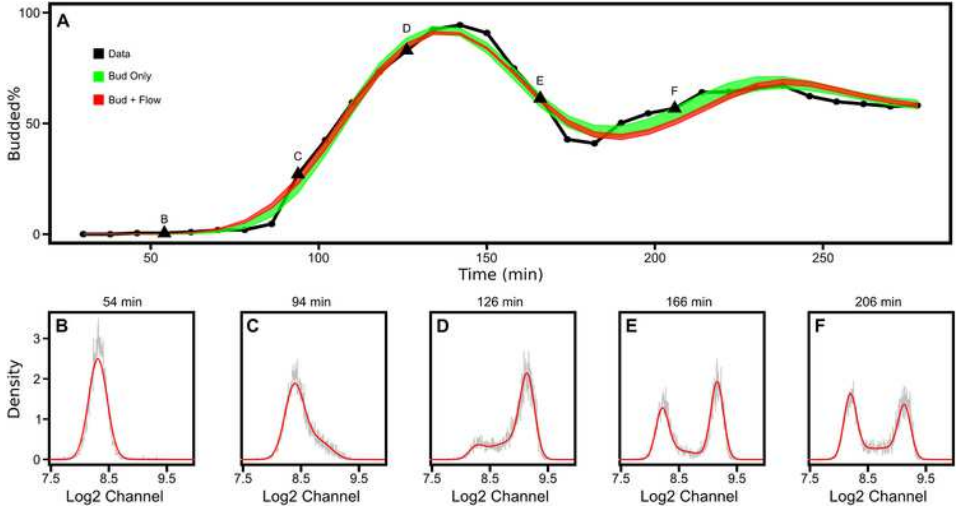}

\caption{Estimated budding and DNA content curves accurately reflect
complex, biologically relevant patterns in the data. \textup{A}:
plot of observed budding index curve (black) and 95\% pointwise
interval estimates from budding index only analysis (green) and
budding index/DNA content analysis (red). \textup{B}--\textup{F}: DNA content
densities (gray) and their posterior mean estimates (red) at five
points in time, highlighting the population's transition from G1
(\textup{B}) through the S phase to G2/M (\textup{C} and \textup{D})
and the effect of its growing asynchrony (\textup{E} and
\textup{F}). The corresponding time points are labeled above the
budding index curve. In all cases, the G1 and G2/M modes are
accurately scaled and located, as is the shape of the distributions
between the modes.}\label{fig:FlowFits}
\end{figure}

Figure~\ref{fig:FlowFits}A is a plot of the observed budding index
curve (black) overlayed with 95\% pointwise interval estimates from
the analysis of only the budding index data (green) and of both
the budding index and DNA content data (red). The latter analysis
estimates the recovery period (Gr) to be slightly shorter and more
variable and estimates cell cycle length to be longer and less
variable than estimated with the budding index data alone. This
is evident in the red confidence bands positioned to the left of the
green between 70 and 100 minutes and to the right of the green between
190 and 225 minutes experimental time. Note that both curves increase
more smoothly and sooner than the observed budding index following
recovery from synchronization. This is likely due to our choice of
the normal distribution to characterize time spent in Gr. It appears
that a left skewed distribution may give a better fit to this feature
in the data.

Figures~\ref{fig:FlowFits}B--F plot observed DNA content densities (gray)
and their posterior mean estimates (red) at five experimental time
points selected to highlight the population's transition from G1 (B)
through the S phase to G2/M (C and D) and the effect of its growing
asynchrony (E and F). The corresponding time points are identified by
labels on the budding index curve Figure~\ref{fig:FlowFits}A. The
observed DNA content densities are discrete and unsmoothed. They are
calculated by normalizing the raw DNA content channel counts and
transforming them, via the change of variables formula, to the log$_2$
scale. The estimates are extremely good: in all cases, the G1 and
G2/M modes are accurately scaled and located and capture the shape of
the distributions between the modes, suggesting that the model is
accurately accounting for the cells transiting the S phase.

%

\subsection{Model evaluation}\label{subsec:eval}
In what follows, we
estimate Bayes factors (BFs) [Kass and Raftery (\citeyear{kassraftery95})] for a series of
pairs of models nested under the fully parametrized CLOCCS model using
importance sampling. These quantities allow us to measure the weight
of evidence in the budding index data in favor of alternate
parametrizations of the model, including variants that drop the
daughter offset and/or one or both parameters of the starting position
distribution. The hierarchy of models we examine is not complete but
accounts for all reasonable alternatives to the full model. The
simplest model, where we set $\mu_0=0$, $\sigma^2_0=0$ and $\delta=0$,
corresponds to a branching process version of the
Bar-Joseph et~al. (\citeyear{barjoseph2004dcc}) model. We employed a separate sampler to
estimate each marginal likelihood and used 100 degrees-of-freedom
multivariate $t$ densities as the importance densities, each with mean
and covariance matrix matching that estimated from a Markov chain
Monte Carlo analysis of the associated model. For purposes of this
calculation, we used only the budding index data to inform the
model and drew 10,000 importance samples for each calculation. The
variance of the normalized weights was less than 1.45 in all cases.
Hence, the effective sample size~[Liu (\citeyear{liu01})] for estimating the
marginal likelihood was never smaller than 4000.

%
\begin{table}[b]
\tabcolsep=0pt
\caption{Estimates of log Bayes
factors (lBFs) for various nested model comparisons given the
budding index data alone. The model indexed by an entry's column is
the larger of the models and is represented in the numerator of
the lBFs in that column; the model indexed by an entry's row is
the smaller of the two. The last two rows of the table provide
the average and standard deviation of the RMSE of the model's
fitted values to the observed budding index data over a
sample of 1000 draws from the posterior}\label{tab:BFs815}
\begin{tabular*}{\textwidth}{@{\extracolsep{\fill}}ld{3.2}d{2.2}d{3.2}d{3.2}d{2.2}d{2.2}d{3.2}c@{}}
\hline
&&&&&
\multicolumn{1}{p{27pt}}{\vspace*{-11pt}\begin{eqnarray*}
\bolds{\mu_0} &\bolds{=}& \bolds{0}\\
\bolds{\delta} &\bolds{=}& \bolds{0}
\end{eqnarray*}}
&
\multicolumn{1}{p{27pt}}{\vspace*{-11pt}\begin{eqnarray*}
\bolds{\mu_0} &\bolds{=}& \bolds{0}\\
\bolds{\sigma^2_0} &\bolds{=}& \bolds{0}
\end{eqnarray*}}
&
\multicolumn{1}{p{27pt}}{\vspace*{-11pt}\begin{eqnarray*}
\bolds{\delta} &\bolds{=}& \bolds{0}\\
\bolds{\sigma^2_0} &\bolds{=}& \bolds{0}
\end{eqnarray*}}
&
\multicolumn{1}{p{27pt}}{\vspace*{-28pt}\begin{eqnarray*}
\hspace*{2pt}\bolds{\mu_0} &\bolds{=}& \bolds{0}\\
\bolds{\delta} &\bolds{=}& \bolds{0}\\
\bolds{\sigma^2_0} &\bolds{=}& \bolds{0}
\end{eqnarray*}}\vspace*{-29pt}
\\
\textbf{Submodel}& \multicolumn{1}{c}{\textbf{Full model}}
& \multicolumn{1}{c}{$\bolds{\mu_0 = 0}$}
& \multicolumn{1}{c}{$\bolds{\delta = 0}$}
& \multicolumn{1}{c}{$\bolds{\sigma^2_0 = 0}$}&
\\ \hline
$\mu_0=0$ & 368.62& 0.00& & & & & & \\
$\delta=0$ & 18.75& &0.00& & & & & \\
$\sigma^2_0=0$ & 31.00& & & 0.00& & & & \\
$\mu_0=\delta=0$ & 364.22& -4.40& 345.47& & 0.00& & & \\
$\mu_0=\sigma^2_0=0$ & 363.78& -4.84& & 332.78& & 0.00& & \\
$\delta=\sigma^2_0=0$& 31.04& & 12.28& 0.04& & & 0.00& \\
$\mu_0=\delta=\sigma^2_0=0$& 359.37& -9.25& 340.62& 328.37& -4.85&-4.41& 328.33&\phantom{0}0.00\\[6pt]
E(RMSE)& 3.70& 13.69& 4.35& 3.99& 13.67& 13.65& 3.92& 13.63\\
$\operatorname{SD}(\mathrm{RMSE})$& 0.16& 0.10& 0.19& 0.15& 0.10& 0.09& 0.14& \phantom{0}0.09\\
\hline
\end{tabular*}
\end{table}

Table~\ref{tab:BFs815} reports estimates of log$_e$ Bayes factors
(lBFs) for various nested model comparisons given the budding
index data. In these tables, the model indexed by an entry's
column is the larger of the models and is represented in the numerator
of the lBFs in that column; the model indexed by an entry's row is the
smaller of the two. As a guide to interpreting these numbers,
Kass and Raftery (\citeyear{kassraftery95}) classify lBFs between 0 and 1 as ``not worth
more than a bare mention,'' those from 1 to 3 ``positive,'' those from 3
to 5 ``strong'' and those greater than 5 ``very strong.'' Using this
scale as a guide, the full CLOCCS model is very strongly preferred to
all alternatives, including the model of~Bar-Joseph et~al. (\citeyear{barjoseph2004dcc}).
The worst alternative sets only $\mu_0=0$. When $\mu_0$ is
constrained to be zero, better fits to the data are achieved by
setting one or the other, or preferably both, of $\delta$ and
$\sigma^2_0$ to zero.

\begin{figure}

\includegraphics{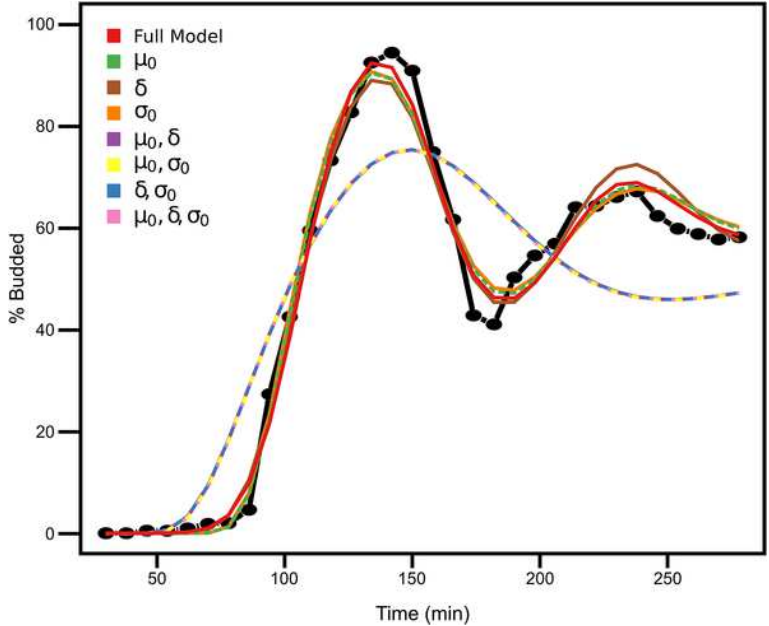}

\caption{Plot of observed budding index curve (black) and posterior
mean fitted curves under each of the competing models for the
budding index data. The full model is plotted in red; the competing
models are obtained by constraining the parameter(s) indicated in
the figure legend to be zero. Quantitative summaries of these fits
can be found in Table~\protect\ref{tab:BFs815}.}\label{fig:SubModelFits}
\end{figure}

Figure~\ref{fig:SubModelFits} depicts posterior mean fits to the
budding index data under each of the competing models. We
estimated the posterior means using the MCMC output that was used to
determine the importance distributions. Each MCMC analysis followed
the same procedure, described in Section~\ref{subsec:fits}, used for
the primary analyses. Note that the fits achieved by all model
variants that set $\mu_0=0$ are visually indistinguishable and
markedly inferior to any variant that allows $\mu_0>0$. The last two
rows of Table~\ref{tab:BFs815} provide estimates of the root mean squared
error (RMSE) of the fits to the budding index data achieved by each
model's posterior mean curve. These estimates reinforce what is
evident from the marginal likelihood and graphical analyses, namely,
that models that do not allow for a nonzero location in the
distribution of initial cell position are markedly inferior to those
that do and that accounting for a mother/daughter offset is
particularly important, at least in the case where the cell population
was arrested using centrifugal elutriation. Finally, these results
demonstrate that the extremely good fits depicted in
Figure~\ref{fig:FlowFits} are the result of a parsimoniously
parametrized model and not due to over-fitting.

%

\section{Discussion}\label{sec:disc}
Synchrony/time-series experiments on populations of cells are essential
for understanding the dynamic processes associated with the cell
cycle. In this paper we have described the CLOCCS model, sampling
models for fitting this model to both budding index and DNA content
data, and a detailed model evaluation. We have demonstrated that
accurate model fits can be obtained using budding index, DNA content
data or both. While previous models only account for one source of
asynchrony, namely, variation cell cycle length
[Bar-Joseph et~al. (\citeyear{barjoseph2004dcc}), Chiorino et~al. (\citeyear{chiorino2001drc}),
Larsson et~al. (\citeyear{larsson2008evs}), Liou, Srienc and Fredrickson
(\citeyear{liou1997spb})],
the CLOCCS model adds two novel sources of asynchrony. These are
variation in initial synchrony and variation due to asymmetric cell
division. In Section~\ref{subsec:eval} we showed that the CLOCCS model
is very strongly preferred to all nested alternatives, including a
branching process version of the model of~Bar-Joseph et~al. (\citeyear{barjoseph2004dcc}).

The more accurate description of population dynamics achieved by the\break
CLOCCS model will allow more accurate deconvolution of dynamic
measurements such as transcript abundance. Additionally, because the
model maps time-series data onto a common cell cycle lifeline,
different data types (e.g., mRNA levels, protein levels, protein
localization, etc.) from multiple synchrony/time-series experiments
can be aligned such that the dynamics of multiple events can be
temporally compared. Furthermore, DNA content measurements are commonly
used to measure cell cycle position in organisms from yeast to
mammals. Thus, the model permits the alignment and comparison of
dynamics of cell cycle events across species, potentially providing an
accurate view of evolutionary changes in cell cycle progression and
regulation.

The model's parameter estimates are also interpretable in terms of
biological quantities associated with the cell cycle, so their
estimates are of independent interest. For example, the measure of
initial synchrony, $\sigma_0$, can be used to tune synchrony protocols
for optimal results. When using budding index data, $\lambda$ and
$\beta$ allow researchers to map temporal events to pre- or post-G1
cell cycle phases. When DNA content data is used, this resolution is
increased and events can be placed accurately into the G1, S or G2/M
phases of the cell cycle.

The CLOCCS model is unique, to our knowledge, for providing a closed
form expression for the likelihood function in a complex branching
process. This expression is written by enumerating and then
marginalizing over the distinct cohorts present in the population at a
given time. The explicit accounting of cohorts allows for extensions
of the model that introduce cohort dependent effects such as one-time
events and effects, such as the mother--daughter offset, that may
diminish with generation. The approach we describe is very general
and has the potential to provide a flexible and efficient alternative
in a range of problems where population balance or branching process
models are used to describe the short term dynamics of a branching
population.

While CLOCCS is better than its nested alternatives, the model can be
improved to better fit experimental data and to better reflect
biological reality. First, our data suggest that a left skewed
distribution with finite support may be more realistic a choice for
the initial position. Second, while our data do not contradict a linear
accumulation of DNA during the S phase, others have suggested alternative
parametrizations [Larsson et~al. (\citeyear{larsson2008evs}), Niemist{\"{o}} et~al. (\citeyear{niemisto2007cme})]. We are
currently exploring a flexibly parametrized S phase function that will
allow inference on its functional form and, by doing so, address a
question of fundamental interest to the greater biological community.
Third, we plan to generalize the model to allow for an unspecified
correlation between mother and daughter cell velocities; this
parameter is currently set to one. Finally, we assume that the delay
due to asymmetric cell division ($\delta$) is constant over time.
Evidence exists, however, that the magnitude of this effect may
change as the experiment progresses. This issue can be addressed
with a suitably parametrized cohort-specific delay term, although the
duration of a typical time-course experiment may limit power to detect
this effect.

The strength of the CLOCCS modeling framework lies in its flexibility.
It is adaptable to new experimental measurements, and given its
ability to use DNA content data, is already applicable to virtually all
biological systems where synchronized populations are studied, most
notably human cell-culture systems. Further integration of the model
with deconvolution and alignment algorithms will provide researchers
with a powerful new tool to aid in the study of dynamic processes
during the cell division cycle. Software implementing the CLOCCS model
can be found at \url{http://www.cs.duke.edu/\textasciitilde amink/software/cloccs}.

\section*{Acknowledgments}
The authors wish to thank Charles Lin for his role in collecting the
data analyzed herein, Allister Bernard for his helpful comments on the
CLOCCS model, members of the Duke Center for Systems Biology for their
invaluable comments as we prepared the manuscript and an AOAS editor
and anonymous referee for their insightful comments that have led to
a greatly clarified presentation.

\printaddresses

\end{document}